\documentstyle[aps,preprint,eqsecnum]{revtex}
\begin{document}
\draft \tighten
\newtheorem{lem}{Lemma}
\newtheorem{cor}{Corollary}
\newtheorem{prop}{Proposition}
\newtheorem{thm}{Theorem}
\title{The central singularity in spherical collapse}
\author{Brien C. Nolan\footnote{e-mail: nolanb@ccmail.dcu.ie}}
\address{School of Mathematical Sciences,\\ Dublin City University,\\
Glasnevin, Dublin 9,\\ Ireland.}
\maketitle
\begin{abstract}
The gravitational strength of the central singularity in spherically symmetric
space-times is investigated. Necessary conditions for the singularity to be
gravitationally weak are derived and it is shown that these are violated in
a wide variety of circumstances. These conditions allow conclusions to be drawn
about the nature of the singularity without having to integrate the geodesic 
equations. In particular, any geodesic with a non-zero
amount of angular momentum which impinges on the singularity terminates in a
strong curvature singularity. 
\newline
\end{abstract}

\section{Introduction}
The theorems of Hawking, Penrose and others predict the occurrence of space-time singularities in a variety of interesting physical situations \cite{HE}. The singularities which have received the most attention over the last years are those which occur in gravitational collapse and the initial cosmological singularity. It seems fair to say that our understanding of these singularities remains at a preliminary stage; little is known about generic 4-d collapse and correspondingly, generic inhomogeneous cosmological singularities. However much progress has been made on the understanding of these singularities under certain simplifying assumptions, e.g. the assumption of spherical symmetry for black holes or of homogeneity for cosmological singularities. See for example the reviews of \cite{burko-ori} and \cite{berger}.

It is in this context that we analyse a particular feature of singularities, 
namely their gravitational strength \cite{tipler}, under the simplifying 
assumption of spherical symmetry. This continues the work initiated in 
\cite{ssss}, but here we concentrate on the central singularity. The notion of the gravitational strength of a singularity was first introduced by Ellis and Schmidt \cite{ellis+schmidt} with the aim of distinguishing between singularities which destroy objects impinging upon them and those which do not. A formal mathematical definition was given by Tipler, based on the familiar idea of modelling an object's physical extension using Jacobi fields along its world-line \cite{tipler}.
Thus let $\gamma:[t_0,0)\to M$ (with tangent $k^a$) be an incomplete causal geodesic running into a singularity as the parameter $t\to 0^-$. We define the sets of Jacobi fields $J_{t_1}, t_0\leq t_1<0$ as follows: $J_{t_1}=\{ \xi^a: (i), (ii), (iii)\}$, where 
\[ (i)\quad g_{ab}\xi^a k^b=0;\]
\[ (ii)\quad \xi^a(t_1)=0;\]
\[ (iii)\quad D^2\xi^a +{R_{bcd}}^ak^bk^d\xi^c =0.\]
(Covariant differentiation along $\gamma$ is represented by $D$.) Note that
the elements of $J_{t_1}$ are space-like vector fields, and so we may refer to their norm,
$||{\vec \xi}||=(g_{ab}\xi^a\xi^b)^{1/2}$.
Given any three (for time-like geodesics; two for null) independent elements of $J_{t_1}$, a volume element along $\gamma$ is constructed by taking the exterior product of the corresponding 1-forms, and a volume $V$ by taking the norm of this 3-form. Then the singularity is said to be gravitationally strong if for all such volumes $V$, we have
\[ \liminf_{t\to 0^-} V(t) = 0.\]
According to Tipler's definition, the singularity is said to be gravitationally weak if this condition does not hold. Thus any object with world-line $\gamma$ will inevitably be crushed by a gravitationally strong singularity.  To emphasize the role played by $\gamma$ here, we will refer to {\em geodesics terminating} in strong or weak singularities.

As pointed out recently and independently by Nolan \cite{ssss} and by 
Ori \cite{ori}, this definition ignores some singularities which would 
destroy objects impinging upon them and so needs a brief addendum. 
Firstly, the definition of a strong singularity ignores the case where 
$V$ diverges to infinity in the approach to the singularity. Subject to 
the strong energy condition and Einstein's equation, $V$ is a convex 
function of $t$, and so cannot diverge in a finite amount of parameter 
time \cite{ck}. However, there are situations where the strong energy 
condition is violated while the weak and dominant energy conditions are 
satisfied. In such a case, convexity of $V$ is not guaranteed and so one 
should allow for the possibility of $V$ diverging. Secondly, as pointed 
out by Tipler, the volume form may stretch infinitely in one direction 
while shrinking to zero in another in such a way that its norm $V$ remains 
finite overall. Such a `spaghettifying' effect clearly signals the end of 
an observer's history. (An observer falling radially into the singularity at the 
centre of a Schwarzschild black hole suffers infinite stretching in the 
radial direction and infinite crushing in the tangential directions. 
The net effect on his volume $V$ is that it is crushed to zero. An 
explicit example of a situation where the radial stretching and 
tangential crushing are exactly cancelled when one calculates $V$ 
was given in \cite{ssss}.) Such situations should also be included 
in the definition of strong singularities. This may be done in a 
logical and succinct fashion following Ori \cite{ori}: a singularity 
is said to be {\em deformationally strong} if it is either (i) Tipler 
strong (i.e. strong in the sense of the paragraphs preceding this one) 
or (ii) if for every $t_1$, there exists an element of $J_{t_1}$ which 
has infinite norm in the limit $t\to 0^-$. A singularity is said to be 
{\em deformationally weak} if it is not deformationally strong.

Our aim is to give useful geometric conditions for the occurrence of 
deformationally strong singularities. We focus on a particular class of 
singularities; those which occur at the centre of spherically symmetric 
space-times. Numerous analyses have predicted the occurrence of such 
singularities inside spherical black holes 
\cite{page,gnedins,brady+smith,burko0,burko1,hod-pir,burko2}, 
as a consequence of gravitational collapse 
\cite{eardley+smarr,ori-pir,jj,singh,christodoulou} and in cosmological models \cite{kras}.
We exploit the available symmetry to develop a condition which is necessary for a singularity to be deformationally weak. In conjunction with an existing necessary condition for a singularity to be Tipler-weak \cite{ck}, we find very severe restrictions on the existence of weak central singularities. In fact it will be shown that {\em any non-radial causal geodesic which approaches $r=0$ terminates in a deformationally strong curvature singularity}. For radial causal geodesics and in particular situations (e.g. assuming a particular matter distribution or a space-like singularity), we can show that these restrictions are violated, i.e. the singularity is deformationally strong. In most cases, we present our necessary conditions as inequalities which must be satisfied along geodesics running into the singularity. However, in many cases we can draw conclusions without having to integrate the geodesic equations. Thus strong curvature singularities can be predicted at the level of the cuvature tensor rather than the geodesics themselves.

In the following section, we study causal geodesics and the volume $V$ along them in spherically symmetric space-times. This allows us to give the result referred to above for non-radial geodesics and to present our main result on radial geodesics in the form of necessary conditions for a central singularity to be deformationally weak. Applications are then given and further comments are given in a concluding section. We emphasize throughout the use of invariant quantities.

\section{Geodesics in spherical symmetry}

We write the line element in double null form;
\begin{eqnarray} ds^2 = -2e^{-2f}dudv + r^2d\Omega^2,\label{sslel}\end{eqnarray}
where $f=f(u,v)$, $d\Omega^2$ is the line element of the unit 2-sphere and $r=r(u,v)$ is the radius function of the space-time (which is a geometric invariant). $u,v$ are null coordinates and the form (\ref{sslel}) is invariant up to $v\to v_1(v)$, $u\to u_1(u)$. A singularity will be referred to as {\em central} if it occurs at $r=0$. In the coordinates of (\ref{sslel}), the Ricci tensor has non-vanishing components
\begin{mathletters}\label{eq2}
\begin{eqnarray}
R_{uu}&=&-2r^{-1}(r_{,uu}+2r_{,u}f_{,u})\label{eq2a}\\
R_{vv}&=&-2r^{-1}(r_{,vv}+2r_{,v}f_{,v})\label{eq2b}\\
R_{uv}&=&-2r^{-1}(r_{,uv}-rf_{,uv})\label{eq2c}\\
R_{\theta\theta}&=&\csc^2\theta R_{\phi\phi}=
1+2e^{2f}(r_{,u}r_{,v}+rr_{,uv}).\label{eq2d}
\end{eqnarray}
\end{mathletters}
Insofar as it is possible, we will describe any curvature tensor terms that we encounter using the following invariants; $e^{2f}R_{uu}$, $e^{2f}R_{vv}$, the Misner-Sharp energy \cite{msnote}
\[ E=\frac{r}{2}(1+2e^{2f}r_{,u} r_{,v}),\]
the Newman-Penrose Weyl tensor Coulomb component (calculated on a principal null tetrad)
\[ \Psi_2=\frac{e^{2f}}{3r}(r_{,uv}+rf_{,uv})-\frac{E}{3r^3},\]
and the Ricci scalar $R$. A useful feature of $E$ is that it offers a simple description of the trapped and untrapped regions of a spherically symmetric space-time; the point $x\in M$ lies on a trapped (untrapped, marginally trapped) 2-sphere iff $\chi = 1-2Er^{-1}$ is negative (positive, zero) at $x$ \cite{hayward}.

Given an arbitrary geodesic $\gamma$ in spherical symmetry, the coordinates of the 2-sphere $(\theta,\phi)$ may be chosen such that the motion proceeds in the hypersurface $\theta=\pi/2$. Thus the tangent to an arbitrary causal geodesic may be written as
\begin{eqnarray} {\vec k} = {\dot u}\frac{\partial}{\partial u}+{\dot v}\frac{\partial}{\partial v}+ Lr^{-2}\frac{\partial}{\partial \phi},\label{gvec}\end{eqnarray}
where we have included the conservation of angular momentum, $r^2{\dot \phi} = L=$ constant. The overdot indicates differentiation with respect to the parameter $t$. The remaining geodesic equations are
\begin{mathletters}\label{geq}
\begin{eqnarray} -2e^{-2f}{\dot u}{\dot v}+L^2r^{-2}&=&-\epsilon, \label{geqa}\\ {\ddot
u}-2f_{,u}{\dot u}^2+L^2e^{2f}r^{-3}r_{,v}&=&0, \label{geqb}\\ {\ddot v}-2f_{,v}{\dot v}^2+L^2e^{2f}r^{-3}r_{,u}&=&0,
\label{geqc} \end{eqnarray}
\end{mathletters}
where $\epsilon=+1$ for time-like geodesics and $\epsilon=0$ for null geodesics.

In a previous paper \cite{ssss}, we studied radial causal geodesics and were able to obtain a useful decomposition of the volume $V$. We found that $V=|axy|$, where $a$ is the norm of a radial Jacobi field and $x,y$ are the norms of two mutually orthogonal tangential Jacobi fields along the geodesic. The key to obtaining this decomposition is the fact that any three such Jacobi fields (which vanish at $t=t_1$) provide a basis for $J_{t_1}$. This complete decomposition of $V$ is not available in the general non-radial ($L\neq 0$) case, but a useful part of it is. This partial decomposition relies on the following facts. 

As we will see, there is always a Jacobi field along $\gamma$ of the form
\begin{eqnarray} {\vec \xi} = x r^{-1}\frac{\partial}{\partial \theta}.\label{jvec}\end{eqnarray}
$x(t)$ must satisfy a certain ODE along $\gamma$; see below.
In the time-like case, the other two elements of a basis for 
$J_{t_1}$ must have non-zero 
components in the 2-space orthogonal to both 
${\vec k}$ and ${\vec \xi}$, and indeed may 
be taken to lie in this 2-space. We consider two such basis elements, 
${\vec \eta}$ and ${\vec \zeta}$.  Now introduce an orthonormal tetrad, 
parallel propagated along $\gamma$. An arbitrary tangent vector ${\vec l}$ 
orthogonal to ${\vec k}$ will only have components on the three spatial 
vectors of the tetrad; let these components be $l^\alpha$, $\alpha=1,2,3$. 
For three such vectors, the corresponding 1-forms satisfy
\[ || {\bf l}_1\wedge{\bf l}_2\wedge{\bf l}_3||=\det\left[l_1^\alpha,l_2^\alpha,l_3^\alpha\right]\] where the columns of the matrix are the tetrad components of the given vectors. For the case where the ${\vec l}_{(i)}$ are elements of $J_{t_1}$, this matrix is a constant matrix multiple (which we will call a transition matrix $T$) of the matrix
\[ \Gamma = \left[ \xi^\alpha, \eta^\alpha, \zeta^\alpha \right].\]
By orthogonality, the volume associated with $\Gamma$ is 
$V_\Gamma=||{\vec \xi}||(||{\bf \eta}\wedge{\bf \zeta}||)$.
In the case of relevance to us, where the ${\vec l}_{(i)}$ are {\em independent} elements of $J_{t_1}$, the transition matrix is non-singular and so we obtain for the $V$ of relevance,
\[ V(t)= \det(T)V_\Gamma.\]
The key point here is that $|x|=||{\vec \xi}||$ appears as a factor of $V(t)$, and so if $x(t)$ is degenerate (i.e. $x\to 0$ or $\infty$) in the limit as the singularity is approached ($t\to 0^-$), then the singularity must be deformationally strong. A similar argument holds in the null case.

We turn to the derivation of the equation satisfied by $x(t)$ in (\ref{jvec}). It is easily verified that the vector field $r^{-1}\partial/\partial\theta$ is parallel propagated along $\gamma$ and has unit norm. Thus (\ref{jvec}) satisfies the geodesic deviation equation iff
\begin{equation} {\ddot x}\delta^a_\theta = -{R_{b\theta d}}^ak^bk^d x.\label{xeqn} \end{equation}
We find that
\begin{eqnarray*} {R_{b\theta d}}^ak^bk^d=
\left( -(2r^{-1}r_{,u}f_{,u}+r^{-1}r_{,uu}){\dot u}^2 
-(2r^{-1}r_{,v}f_{,v}+r^{-1}r_{,vv}){\dot v}^2\right.\nonumber\\
\left.+\sin^2\theta(1+2e^{2f}r_{,u}r_{,v}){\dot\phi}^2-2r^{-1}r_{,uv}{\dot u}{\dot v}
\right)\delta^a_\theta.\nonumber\end{eqnarray*}
This term is controlled solely by $r(t)$. The evolution of $r$ along $\gamma$ is given by ${\dot r}=r_{,u}{\dot u}+r_{,v}{\dot v}$. The second derivative can be worked out and simplified with the use of the geodesic equations, resulting in 
\[{\ddot r}= (2r_{,u}f_{,u}+r_{,uu}){\dot u}^2 +(2r_{,v}f_{,v}+r_{,vv}){\dot v}^2-2e^{2f}rr_{,u}r_{,v}{\dot\phi}^2 + 2r_{,uv}{\dot u}{\dot v}.\]
Comparing the last two equations along $\gamma$ (on which $\theta=\pi/2$), we see that the equation (\ref{xeqn}) becomes
\begin{equation} {\ddot x}+\left(\frac{L^2}{r^4}-\frac{\ddot r}{r}\right)x = 0.\label{maineq}\end{equation}
The existence of the angular momentum term plays a vital role as we will see in the following section. We treat the radial $(L=0)$ and non-radial $(L\neq 0)$ cases separately, dealing with the latter first.

\section{Non-radial geodesics}
We may assume without loss of generality that $L>0$. Defining $y=r^{-1}x$, the equation (\ref{maineq}) may be written in the self adjoint form 
\begin{equation} \frac{d}{dt}(r^2\frac{dy}{dt})+\frac{L^2}{r^2}y=0.\label{yeqn}\end{equation}
We can obtain the asymptotic behaviour of $x$ in the limit as the singularity is approached as follows. First, we move the singularity out to infinite parameter value. The origin and temporal orientation of the affine parameter/proper time $t$ has been fixed so that the singularity at $r=0$ is approached as $t\to 0^-$. We define $s=-t^{-1}$, so that the singularity is approached as $s\to +\infty$. Defining $R(s)=r(t)$, where $r(t)$ means $r(u,v)|_{u=u(t),v=v(t)}$, i.e. this indicates the dependence of $r$ on the parameter $t$ in the solution of the geodesic equations, equation (\ref{yeqn}) becomes
\begin{equation} (s^2R^2y^\prime)^\prime+\frac{L^2}{s^2R^2}y=0,\label{yseqn}\end{equation}
where the prime denotes differentiation with respect to $s$. Thus the equation is of the form
\begin{equation} (p(s)y^\prime)^\prime-q(s)y=0.\label{lgeq}  \end{equation}
We can use the Liouville-Green asymptotic formula to obtain the leading order behaviour of $y$ and hence $x$ as $s\to\infty$. We quote in full the following theorem which appears as Theorem 2.2.1 of \cite{eastham}.

\begin{thm}\label{thm1}
Let $p$ and $q$ be nowhere zero and have locally absolutely continuous first derivatives in an interval $[a,\infty)$. Let
\begin{equation} \frac{(pq)^\prime}{pq}=o\left\{(\frac{q}{p})^{1/2}\right\},\qquad (s\to\infty) \label{a}\end{equation}
and let
\begin{equation}
\left(p^{-1/2}q^{-3/2}(pq)^\prime\right)^\prime\in L(a,\infty)\label{b}\end{equation}
Let ${\rm Re}((\frac{q}{p}+w^2)^{1/2})$ have one sign in $[a,\infty)$ where $w=(pq)^\prime/(4pq)$. Then (\ref{lgeq}) has solutions $y_1,y_2$ with asymptotic behaviour
\begin{equation} y_{1,2}\sim (pq)^{-1/4}\exp\left(\pm\int_a^s (\frac{q}{p}+w^2)^{1/2} d{\bar s}\right),\qquad (s\to\infty).\label{c}\end{equation}
\end{thm}

We now turn to the application of this theorem to the equation (\ref{yeqn}). We have here
\[ p=s^2R^2,\qquad q=-\frac{L^2}{s^2R^2}.\]
Recall that we are working under the hypothesis that $R(s)$, satisfying the geodesic equation, approaching $R=0$ as $s\to\infty$. Thus there exists a real $a$ such that neither $p$ nor $q$ are zero in the interval $[a,\infty)$. We have
\[ p^\prime = 2sR^2+2s^2R^\prime = 2t^{-1}r^2+2{\dot r},\]
\[q^\prime =2L^2(s^{-3}R^{-2}+s^{-2}R^{-4}R^\prime)=2L^2(t^3r^{-2}+t^4r^{-4}{\dot r}).\]
The requirement that these functions be locally absolutely continuous on $[a,\infty)$, i.e. that they be absolutely continuous on compact subsets of $[a,\infty)$, is very weak. This would follow from local boundedness of $p^\prime, q^\prime$ which itself will follow from the assumption of the existence of the geodesic on the interval, i.e. from the fact that ${\dot r}$ is defined and bounded away from the singularity. Also,
\[ pq = -L^2,\]
which is constant, and so conditions (\ref{a}) and (\ref{b}) are automatically satisfied. Note also that $w=0$. Thus the hypotheses of the theorem are satisfied and the conclusion (\ref{c}) yields (on removing a constant unimodular factor)
\[ y_{1,2}\sim L^{-1/2}\exp\left(\pm i\int_a^s \frac{L}{{\bar s}^2R^2({\bar s})}\, d{\bar s}\right).\]
Writing this in terms of the affine parameter/proper time $t$, we obtain for the two real independent solutions $x_{1,2}$ of (\ref{maineq})
\begin{eqnarray}
x_1(t)&\sim&L^{-1/2}r(t)\cos\left(\int_\epsilon^t \frac{L}{r^2(t^\prime)}\,dt^\prime\right),\\
x_2(t)&\sim&L^{-1/2}r(t)\sin\left(\int_\epsilon^t \frac{L}{r^2(t^\prime)}\,dt^\prime\right).
\end{eqnarray}
Now let $J^a(t)=x(t)r^{-1}\delta^a_\theta$ be a Jacobi field which vanishes at some arbitrary time $t_1<0$, i.e. $x(t_1)=0$. 
Since {\em both} of the independent solutions $x_{1,2}$ of (\ref{maineq}) approach zero as $t\to0^-$, we conclude that the particular linear combination which gives the present $x(t)$ will also approach zero in this limit. Thus we see that in every case, $x(t)\to 0$ in the limit as the singularity is approached.
This proves the following result, which we note is independent of any 
energy conditions.

\begin{prop}\label{prop1}
Let $\gamma$ be a non-radial causal geodesic in a spherically symmetric space-time $(M,g)$. If $\gamma$ runs into the centre $r=0$ in finite parameter time, either in the past or in the future, then $\gamma$ terminates in a deformationally strong curvature singularity.
\end{prop}

Thus any singularity, naked or covered, which is reached by a non-radial 
causal geodesic is deformationally strong. These geodesics have not been 
widely studied and deserve some attention. It would be of interest to 
know, for example, if those space-times which admit radial geodesics 
with past endpoints on a central singularity - i.e. which admit naked 
singularities - also admit non-radial geodesics with the same, or indeed 
if there are non-radial geodesics whose futures terminate at the past 
endpoints of the naked singularity geodesics. If such geodesics exist, 
then these singularities should be considered genuine; their existence 
has a destructive effect on certain observers in the space-time. We hope 
to address the question of the existence of such geodesics in future work. A useful starting point would be the invariant equation for the evolution of $r$ along the geodesic, i.e. equation (\ref{rdd}) below.

\section{Radial geodesics}

In the radial case, the angular momentum term $L$ vanishes and (\ref{maineq}) reads
\begin{equation} r{\ddot x}-{\ddot r}x=0.\end{equation}
The unique solution (modulo an irrelevant constant factor) of this equation satisfying $x(t_1)=0$ is 
\begin{eqnarray} x(t)= r(t)\int_{t_1}^t \frac{ds}{r^2(s)}. \label{xsol}\end{eqnarray}
Notice then that if the singularity is non-central, $x(0^-)$ is non-zero and finite. For a central singularity, we have the following \cite{ssss}. 

\begin{lem}\label{lem1}
Let
\[ I =\int_{t_1}^0 \frac{ds}{r^2(s)}.\]
(i) If the integral $I$ converges, then $x\to 0$ as $t\to 0^-$.
\newline
(ii) If $I$ diverges, then
\[ \lim_{t\to 0^-} x(t) = \lim_{t\to 0^-}-\frac{1}{\dot r}(t).\]
\end{lem}

This is a straight application of l'Hopital's rule. From this and the comments above, we have the following useful corollaries.

\begin{cor}\label{cor1}
If a radial causal geodesic $\gamma$ terminates in a deformationally weak central singularity, then along $\gamma$, $\lim_{t\to 0^-}{\dot r}(t)$ is non-zero and finite. 
\end{cor}

\begin{cor}\label{cor2}
Let the conditions of corollary \ref{cor1} be satisfied. Then
there exists $c_0>0$ such that \[ r(t) \sim c_0|t|\qquad {\rm as} \qquad t\to 0^-.\]
\end{cor}

This yields a useful necessary condition for deformational weakness of the singularities under consideration.
We use corollary \ref{cor1} above and some established results on weakness of singularities \cite{ck} to derive a new necessary condition for deformational weakness of central singularities reached by radial geodesics. This condition seems unlikely to be satisfied in many circumstances. We require the following preliminary basic results.

\begin{lem}\label{lem2}
Let $\alpha\in C(0,b]$ for some $b>0$. Suppose that 
\[ \int_0^r \alpha(s) ds\]
converges for all $r\in(0,\epsilon) (b>\epsilon>0)$. Then $\lim_{r\to 0}r\alpha(r)=0$.
\end{lem}

Proof: Let $r$ be fixed and define $\beta$ on $[0,r]$ by
\[ \beta(x)=\int_x^r \alpha(s)\,ds.\]
Then $\beta\in C[0,r]\cap C^1(0,r]$. By Taylor's theorem, for every $x\in[0,r]$, there is an $r_1\in(x,r)$ such that
\begin{eqnarray*}
\beta(x)&=&\beta(r)+\beta^\prime(r_1)(x-r)\\
&=& (r-x)\alpha(r_1).\end{eqnarray*}
In particular,
\[ H(r):=\int_0^r \alpha(s)\,ds =\beta(0) = r\alpha(r_*)\]
for some $r_*\in(0,r)$.
By hypothesis, $H(r)$ exists and is finite for all sufficiently small $r$. Thus
\begin{eqnarray*}
0&=&\lim_{r\to 0}|H(r)|\\
&=&\lim_{r\to 0}|r\alpha(r_*)|\\
&\geq&\lim_{r\to 0}|r_*\alpha(r_*)|\\
&=&\lim_{r_*\to 0}|r_*\alpha(r_*)|\geq 0.
\end{eqnarray*}
Replacing $r_*$ by $r$ in the last line gives the required result.

\begin{lem}\label{lem3}
Let $r(t)$ satisfy the differential equation
\[ {\ddot r}=\alpha(r) \]
on $(0,b]$ where $\alpha\in C(0,b]$ for some $b>0$.
Suppose that $\lim_{t\to 0} r(t)=0$ and that $\lim_{t\to 0}{\dot r}$ exists and is finite. Then $\lim_{r\to 0} r\alpha(r) =0$.
\end{lem}
Proof: For sufficiently small $r$ and for $t>0$, we may integrate the differential equation to obtain
\[ {\dot r}^2(t) =2\int_{r_0}^r \alpha(r^\prime)\,dr^\prime +{\dot r}^2(t_0),\]
where $r(t_0)=r_0\leq b$. Then by hypotheses,
\begin{eqnarray*}
\int_{r_0}^0 \alpha(s)\,ds &=& \lim_{r\to 0}\int_{r_0}^r \alpha(r^\prime)\,dr^\prime\\
&=& \frac{1}{2}\lim_{t\to 0} ({\dot r}^2(t) - {\dot r}^2(t_0))
\end{eqnarray*}
exists and is finite. Now apply lemma \ref{lem2} to obtain the result.

Next, we recall a result of Clarke and Krolak (a direct consequence of their corollary 3 \cite{ck}). For this we note that if a singularity is deformationally weak, then it is Tipler-weak. Note that this result assumes the strong energy condition and Einstein's equation (or equivalently, the time-like and null convergence conditions).

\begin{lem}\label{lem4}
If a causal geodesic $\gamma$ terminates in a deformationally weak singularity, then along $\gamma$, 
\[ \lim_{t\to 0^-} t^2 R_{44} =0,\]
where $R_{44}:=R_{ab}k^ak^b$.
\end{lem}

In order to use lemma \ref{lem3}, we note the following. 
We have ${\dot r} = r_{,u}{\dot u}+r_{,v}{\dot v}$. Differentiating again, 
using the geodesic equations (\ref{geq}) and grouping terms appropriately, we get the following form for ${\ddot r}$ (for generality, we include angular momentum in the following expression):
\begin{eqnarray} r^{-1}{\ddot r}&=&-\frac{1}{2}R_{44}+\epsilon(\frac{E}{r^3}+2\Psi_2-\frac{R}{12})+L^2r^{-2}(\frac{1}{r^2}+3\Psi_2).\label{rdd}\end{eqnarray}
The right hand side is to be viewed as a function of the parameter $t$ 
(i.e. it is assumed that the geodesic equations are solved), which is 
smooth for $0<|t|<\delta$ for some $\delta$ and singular at $t=0$. 
(The degree of smoothness is not particularly significant; continuity 
is sufficient. However there is very little restriction in assuming a 
higher degree of differentiability for $|t|>0$; we are interested in 
singularities occurring at $t=0$.) In the present situation, corollary 
\ref{cor2} applies and so we can use the inverse function theorem to 
write the right hand side of (\ref{rdd}) as a function of $r$ which is 
continuous on $(0,b]$ for some positive $b$. Thus lemma \ref{lem3} applies.

Combining lemma \ref{lem4} with corollaries \ref{cor1} and \ref{cor2}, we obtain our main result.

\begin{prop}\label{prop2}
If a radial causal geodesic $\gamma$ terminates in a deformationally weak central singularity, then along $\gamma$
\[ \lim_{r\to 0}r^2R_{44} = \lim_{r\to 0} r{\ddot r} = 0.\]
\end{prop}

The usefulness of the result comes from the fact that on the one hand 
we have two independent conditions which must be satisfied by weak 
central singularities, and on the other, the parameter $t$ does not 
appear explicitly in the relevant quantities. These features are 
emphasized by using (\ref{rdd}) and the following equation for $R_{44}$:
\begin{eqnarray} R_{44}&=&R_{uu}{\dot u}^2+R_{vv}{\dot v}^2+2\epsilon(\frac{E}{r^3}+\Psi_2-\frac{R}{6})+4L^2r^{-2}(\frac{E}{r^3}+\Psi_2-\frac{R}{24}).\label{r44}\end{eqnarray}

An immediate consequence of equations (\ref{r44}), (\ref{rdd}) and proposition \ref{prop2} 
is the following:
\begin{cor}\label{cor3}
If a radial causal geodesic $\gamma$ terminates in a deformationally weak central singularity, then along $\gamma$
\begin{eqnarray} \lim_{r\to 0} \epsilon r^2(\frac{E}{r^3}+2\Psi_2-\frac{R}{12})=0.\label{term}\end{eqnarray}
\end{cor}
Notice that this result is vacuous for radial null geodesics; this is a consequence of the fact that these are principal null directions in spherical symmetry. On the other hand, we see that the strength of the singularity approached by a radial null geodesic is completely controlled by the behaviour of $R_{44}$ in the limit as the singularity is approached. Indeed we can give the following result.
\begin{cor}\label{cor4}
A radial null geodesic which reaches $r=0$ in finite parameter time terminates in a deformationally weak singularity if and only if there exists $\epsilon>0$ such that $rR_{44}$ is integrable as a function of $r$ on $[0,\epsilon]$.
\end{cor}
Proof: In the radial case, the volume element along the geodesic is a constant multiple of $x^2$, where $x$ satisfies (\ref{xsol}). This follows from the fact that the second Jacobi field may be chosen to be $y(t)r^{-1}\csc\theta\delta^a_\phi$, with $y$ also satisfying (\ref{xsol}). See \cite{ssss}. Thus by Lemma 1, the singularity is deformationally weak if and only if ${\dot r}$ is finite in the limit as the singularity is approached. 
But using (\ref{rdd}), we can write
\[ {\dot r}^2=-\int_{r_0}^r r^\prime R_{44}(r^\prime)\, dr^\prime + {\dot r}^2(t_0),\]
where $r_0=r(t_0)$. The result follows immediately from this integral.

This result shows that the converse of Lemma 4 (with $t$ replaced by $r$) is true for radial null
geodesics. 
This may be useful as this lemma has been used widely 
in studies of radial null geodesics emanating from central 
singularities; see e.g. \cite{ori-pir,jj}. We now have the useful converse, that if the condition used to demonstrate deformational strength of the singularity fails, then the singularity is necessarily deformationally weak.

In the following section, we analyze the conditions $r^2R_{44}\to0$ and 
$r{\ddot r}\to 0$ subject to various assumptions. In several cases, we show how deformationally strong central singularities may be identified without having to integrate the geodesic equations. 

\section{Applications}

In this section, we assume the following situation obtains: there exists a radial time-like geodesic $\gamma$ which runs into $r=0$ in a finite amount of parameter time. Then the origin of the parameter $t$ along the geodesic may be translated so that $r(0)=0$. We set aside the issue of the existence of such geodesics.

We assume that the dominant and strong energy conditions are satisfied by the energy-momentum tensor of the space-time and that Einstein's equation holds. The dominant energy condition states that ${T^a}_bl^b$ is past-directed and causal for any future-directed time-like $l^a$. Then in particular $T_{ab}l^al^b\geq 0$ for all causal $l^a$. We choose units so that Einstein's equation is $G_{ab}=8\pi T_{ab}$. Given this equation, the strong energy condition is equivalent to $R_{ab}l^al^b\geq 0$ for all causal $l^a$.
In particular, the dominant energy condition yields the following inequalities:
\begin{eqnarray} R_{uu}\geq 0,\qquad R_{vv}&\geq& 0,\nonumber\\
z^2:=\frac{E}{r^3}+\Psi_2+\frac{R}{12}&\geq& 0.\nonumber\end{eqnarray}

The approach we take here is to derive general results about the strength of the singularity without integrating either field equations or geodesic equations. We will try to derive general results based on certain geometric assumptions, and otherwise, restrict to particular matter models. In this case, we will focus on two important and widely studied cases; a scalar field in a source free electric field, and a perfect fluid. In the former case, the Ricci tensor is given by
\begin{eqnarray} R_{ab}=2\nabla_a\phi\nabla_b\phi + \frac{Q^2}{r^4}E_{ab},\label{ric4phi}\end{eqnarray}
where $\phi$ is the scalar field, $Q$ is the constant electric charge, $E_{ab}=-g_{ab}+2r^2s_{ab}$ and $s_{ab}$ is the standard metric on the unit 2-sphere. We have $R=2g^{ab}\nabla_a\phi\nabla_b\phi$ and
\begin{eqnarray} \frac{E}{r^3}+\Psi_2+\frac{R}{12}&=&\frac{Q^2}{2r^4}.\label{z4phi}\end{eqnarray}
We note that the dominant and strong energy conditions are automatically satisfied by this matter distribution. THe neutral case can be studied by setting $Q=0$. Burko \cite{burko2} has shown that under the assumption of spatial homogeneity, the central singularity in this model is deformationally 
strong.

For a perfect fluid with flow vector $u^a$, energy density $\rho$ and pressure $p$, we have
\begin{eqnarray} (8\pi)^{-1}R_{ab}&=&(\rho+p)u_au_b+\frac{1}{2}(\rho-p)g_{ab},\label{ric4pf}\\
R&=&8\pi(\rho-3p),\\ 
\frac{E}{r^3}+\Psi_2 &=&\frac{4\pi\rho}{3}.\label{z4pf}\end{eqnarray}
The dominant energy condition requires $\rho + p\geq0$, $\rho-p\geq 0$.

We now proceed to investigate the consequences of proposition \ref{prop2} and corollary \ref{cor3} under the following cases.
Unless otherwise stated, we make no assumptions about the matter distribution, other than that the dominant and strong energy conditions are satisfied. Throughout the remainder of this section, asymptotic relations, limiting values etc. refer to the limit as $r\to 0$ along a geodesic which terminates in a deformationally weak singularity.

For radial time-like geodesics, $L=0,\epsilon=1$ and so the conclusion of corollary \ref{cor3} is that, along a radial geodesic terminating in a deformationally weak central singularity,
\begin{eqnarray} r^2(\frac{E}{r^3}+2\Psi_2-\frac{R}{12}) &\to& 0.\label{radcon}\end{eqnarray}
 For convenience, we give
\begin{eqnarray*}
R_{44}&=&R_{uu}{\dot u}^2+R_{vv}{\dot v}^2 + 2(\frac{E}{r^3}+\Psi_2-\frac{R}{6}) ,\\
r^{-1}{\ddot r}&=&-\frac{1}{2}R_{44}+(\frac{E}{r^3}+2\Psi_2-\frac{R}{12}).\end{eqnarray*}

It is difficult to make any general statements without making further assumptions. However (\ref{radcon}) may prove to be a useful condition to use to check the strength of certain singularities. We can make some progress in the case where
\begin{eqnarray} \frac{E}{r^3}+\Psi_2-\frac{R}{6} \geq 0.\label{c2}\end{eqnarray} Then in addition to 
(\ref{radcon}), we must have
\[ r^2(\frac{E}{r^3}+\Psi_2-\frac{R}{6})\to 0.\]
We can write these as (any two of)
\[ r^2(z^2-\frac{R}{4})\to 0,\qquad r^2(z^2-\frac{E}{r^3})\to 0,\qquad r^2(z^2+3\Psi_2)\to0.\]

If any one of $E\leq 0$ or $R\leq 0$ or $\Psi_2\geq0$ holds, then these give 
\[ r^2\Psi_2,\frac{E}{r},r^2R \to 0\]
as conditions which must hold at a weak singularity. Notice then that the singularity must be untrapped ($1-2E/r\to1>0$ as the singularity is approached). That is:
\begin{cor}\label{cor5}
If a radial time-like geodesic $\gamma$ terminates in a deformationally weak central singularity and if the inequality (\ref{c2}) and at least one of the inequalities
\[ E\leq0,\qquad R\leq0,\qquad \Psi_2\geq 0\]
hold in the limit as the singularity is approached, then the singularity is untrapped. 
\end{cor}

It is worthwhile investigating the consequences of (\ref{c2}) for the two matter models mentioned above. For a scalar field in an electric field, this is equivalent to
\[\frac{Q^2}{r^4}\geq g^{ab}\nabla_a\phi\nabla_b\phi,\]
which may be described as the case where the electric field dominates the scalar field. Then we conclude that
\begin{eqnarray*}
g^{ab}\nabla_a\phi\nabla_b\phi &=& \frac{Q^2}{r^4} +o(r^{-2}),\\
\frac{E}{r}&=&\frac{Q^2}{2r^2} + o(1),\\
\Psi_2&=&-\frac{Q^2}{6r^4}+o(r^{-2}).\end{eqnarray*}
Notice that the singularity must be naked. This should be considered in conjunction with the result of \cite{burko2}, where evidence is
presented that in this case, the singularity must be time-like but that the evolution tends to avoid this situation (i.e. the scalar field dominates). For the case of a neutral scalar field $(Q=0)$, (\ref{c2}) becomes
\[g^{ab}\nabla_a\phi\nabla_b\phi \leq 0.\]
Setting $Q=0$ in the previous trio of equations, we see that in this case the singularity is untrapped.

For a perfect fluid (\ref{c2}) reads
\[ p\geq 0,\]
and for a deformationally weak central singularity subject to this condition, we conclude that
\begin{eqnarray} r^2p\to 0,\qquad 2\pi\rho=\frac{E}{r^3}+o(r^{-2}).\label{pf}\end{eqnarray}

Ori and Piran \cite{ori-pir} considered self-similar collapse of a 
perfect fluid with a barotropic equation of state (which must necessarily 
be of the form $p=\gamma \rho$, $\gamma$ constant). Every such space-time 
includes a central singularity at $r=0, t=0$ where $t$ is an orthogonal 
time coordinate, fixed by demanding that it measures proper time of an 
observer at the regular centre. This point is referred to as the origin. 
It is found that $r^2\rho=D(x)$ where $x=r/t$ is the similarity variable. 
Thus by (\ref{pf}) (which applies if $0\leq k<1/3$), 
a radial time-like geodesic running into the origin 
terminates in a deformationally strong singularity provided 
$\lim D(x)\neq 0$ along the geodesic. The existence of such 
geodesics is readily demonstrated using the results of \cite{ori-pir}. 
The corresponding result for outgoing radial null geodesics was proven by 
Ori and Piran; this includes the interesting case of future-pointing outgoing
radial null geodesics originating at the singularity i.e. the case of a 
naked singularity. The present result shows that these singularities will destroy an observer impinging upon them.
Deformationally strong spacelike singularities have also been detected in 
the gravitational collapse of a perfect fluid (with and without an electric field)
under the assumption of spatial quasi-homogeneity \cite{dori}.

Recall from above that the present case (\ref{c2}) includes dust. The analysis which follows gives a non-trivial example of the demonstration of the deformational strength of a singularity which does not rely on solving the geodesic equations. 
Thus we investigate the latter condition in (\ref{pf}), i.e.
\[ r^2(2\pi\rho - \frac{E}{r^3}) \to 0.\]

For the case of marginally bound dust, we have
\[ 4\pi\rho=\frac{E^\prime}{r^2r^\prime},\]
where the prime denotes differentiation with respect to a spatial coordinate $x$ which labels points in the slices orthogonal to the fluid flow. The Einstein equations in this case yield $E=E(x)$ and
\[ r^3(x,\tau)=\frac{9}{2}E(\tau_0(x)-\tau)^2,\]
where the fluid flow vector is $u_a=-\nabla_a\tau$. See \cite{eardley+smarr,jj} for details. There is freedom in the choice of the coordinate $x$ which allows $x\to X(x)$. This may be utilised by taking $r=x$ on the initial slice $\tau=0$. This choice has the advantage of specifying $\tau_0(x)$:
\[ \tau_0(x)=3\sqrt{\frac{2}{E}}x^{3/2}.\]
The central singularity in this model appears at $\tau=\tau_0(x)$; that singularity for which we also have $x=0$ is of interest for studies of cosmic censorship. Jhingan and Joshi argue that the appropriate form for $E$ is
\[ E(x)=\sum_{n=0}^\infty E_nx^{n+3},\]
where the $E_n$ are constants. ($E$ is not required to be analytic and this is not supposed to be implied by the form above. In what follows, all we require is that $E\sim E_0x^3, x\to 0$ and that this relation is differentiable at $x=0$. This form for $E$ is the most general that ensures a finite, non-singular initial state for the matter.)
With this assumption, a straightforward calculation gives
\[r^2(4\pi\rho-e\frac{E}{r^3})=\frac{EE^\prime(\tau_0-\tau)-4\tau_0^\prime 
E^2}{E^\prime r(\tau_0-\tau)+2\tau_0^\prime Er}.\]
This quantity definitely diverges at a central singularity 
$r=(\tau_0-\tau)=0, x\neq 0$; such a singularity must be deformationally 
strong. In the case $r=(\tau_0-\tau)=x=0$, we have
\[ r^2(4\pi\rho -2\frac{E}{r^3})\sim
\frac{3E_0x^2+\frac{4}{3\sqrt{2}}E_0^{7/2}E_1^{-1}}{3^{5/3}2^{-1/3}E_0^{1/3}(\tau_0-\tau)^{5/3}-2^{-1/6}3^{-1/3}E_0^{17/6}E_1^{-1}x(\tau_0-\tau)^{2/3}}.\]
Again, this quantity (generically) diverges, giving a strong curvature singularity.
We emphasise that it was not necessary to integrate the geodesic equations in order to reach this conclusion.

\section{Conclusions}
We have exploited the symmetry properties of causal geodesics in 
spherically symmetric space-times to study the effect of a central 
singularity on an observer who impinges upon it. We have been able 
to demonstrate the destructive effect (deformational strength) of 
the singularity in several cases, and have given very finely tuned 
conditions which must hold in order that this destruction need not occur. 
In particular, any non-radial geodesic approaching a central singularity must terminate in a deformationally strong central singularity. This may be understood as follows. The focussing term $R_{44}$ will include the term $R_{\phi\phi}{\dot \phi}^2$, which by the conservation of angular momentum equals $L^2R_{\phi\phi}r^{-4}$. This introduces strong curvature along the geodesic which contributes to the destructive effect. Of course this does not include the vacuum case, and so we conclude that the angular momentum must also cause a significant amount of shear to develop along the geodesic which, via the Raychaudhuri equation, contributes to the strong focussing effect. 

The main advantage of our approach was that it did not require the 
integration of the geodesic equations; it was possible to predict 
the deformational strength (or weakness) of certain singularities 
by calculating the Riemann tensor rather than its tetrad components 
\cite{ck}. Of course one needs to address the issue of the existence 
of geodesics which run into the singularity (i.e. the question of the 
{\em existence} of the singularity) in the situations studied above. 
However this can often be done {\em without} having to obtain the detailed 
and subtle information required to apply the results of \cite{ck} (see, for example, \cite{burko3} for a thorough application of these results to the null weak Cauchy horizon singularity in spherical black holes). 
For example, if $z^2=0$ is satisfied along a causal geodesic - as is the case for a neutral scalar field - then (\ref{rdd}) reads
\[r{\ddot r}=-\frac{r^2}{2}(R_{uu}{\dot u}^2+R_{vv}{\dot v}^2)-\epsilon \frac{E}{r}+\frac{L^2}{r^2}(1-3\frac{E}{r}).\]
In a trapped region, $-E/r < -1/2$, and so ${\ddot r}<0$. Assuming the absence of singularities away from $r=0$, this is sufficient to ensure that the geodesic runs into the centre. If $L\neq 0$, this will be a deformationally strong singularity.

The analysis here was made possible by the assumption of spherical symmetry. One would expect similar results in space-times with hyperbolic and plane symmetry. It may also be possible to extend the applicability of the idea of determining the nature of singularities from simple geometric quantities to more general situations, e.g. axially symmetric space-times or homogeneous cosmologies. We note that in this vein, significant progress has been made recently on the issue of the connection between a well-behaved metric and weak singularities \cite{ori2}.

\section*{Acknowledgements}
I am grateful to Amos Ori for helpful discussions and to the referee of the original version of this paper for useful comments.

\end{document}